\documentclass[shortnote,twocolumn,pdftex]{jpsj3}
\usepackage{txfonts}
\usepackage{amsmath}
\usepackage{bm}
\usepackage{braket}

\newcommand\HM{\hat{\mathcal{H}}_\gamma}
\newcommand\HL{\hat{\mathcal{H}}_\mathrm{L}}
\newcommand\HR{\hat{\mathcal{H}}_\mathrm{R}}
\newcommand\HI{\hat{\mathcal{H}}_\mathrm{I}}
\newcommand\I{\mathrm{i}}
\newcommand\D{\mathrm{d}}
\newcommand\E{\mathrm{e}}
\newcommand\para{{\uparrow\!\uparrow}}
\newcommand\apara{{\uparrow\!\downarrow}}

\def\Xint#1{\mathchoice
   {\XXint\displaystyle\textstyle{#1}}%
   {\XXint\textstyle\scriptstyle{#1}}%
   {\XXint\scriptstyle\scriptscriptstyle{#1}}%
   {\XXint\scriptscriptstyle\scriptscriptstyle{#1}}%
   \!\int}
\def\XXint#1#2#3{{\setbox0=\hbox{$#1{#2#3}{\int}$}
     \vcenter{\hbox{$#2#3$}}\kern-.5\wd0}}

\def\dashint{\Xint-}

\title{Tunable Spin Seebeck Diode with Magnonic Spin Tunneling Junction}

\author{Daisuke Miura\thanks{dmiura@solid.apph.tohoku.ac.jp} and Akimasa Sakuma}
\inst{Department of Applied Physics, Tohoku University, Sendai 980-8579, Japan}

\abst{We theoretically investigate the spin--wave spin current induced by the spin Seebeck effect in magnonic spin tunneling junctions (MSTJs) for arbitrary magnetization directions.
We show that the MSTJ functions as a \textit{tunable} spin Seebeck diode
in which
the tunneling spin current can be turned on and off with high efficiency by controlling the magnetization direction.
}

\begin{document}
\maketitle

\begin{figure}[tb]
\centering
\includegraphics[width=0.4\textwidth]{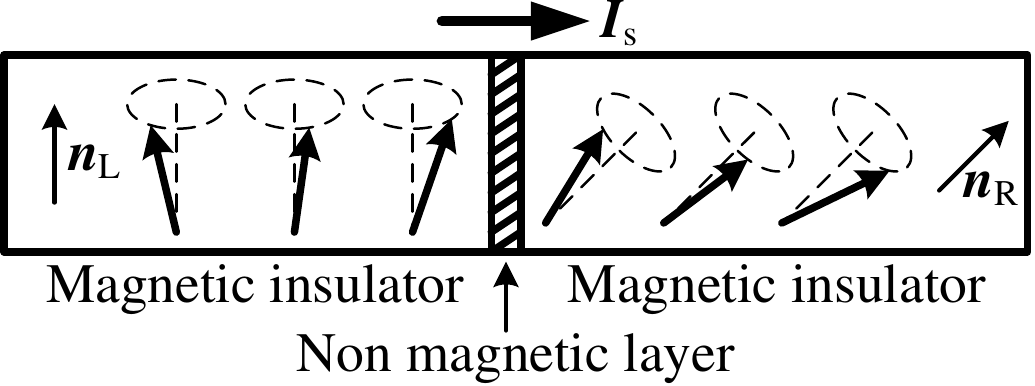}
\caption{Schematic view of the magnonic spin tunneling junction.}
\label{fig:fig1}
\end{figure}

In magnetic insulators,
the spin--wave spin current (magnon current) is a pure spin current without charge flow,
and therefore, all magnonic computers fabricated from magnetic insulators
are expected to operate without Joule heat loss.\cite{Chumak2014}
As a candidate material,
yttrium iron garnet (YIG) potentially has a long propagation length for spin waves;\cite{Schneider2008,Kajiwara2010}
thus, magnon transport in YIG
has attracted considerable attention
in the fields of spintronics,\cite{Zutic2004} magnonics,\cite{Lenk2011} and spin caloritronics.\cite{Bauer2012}
Temperature gradient is one of the forces directly driving the magnon current,
and this phenomenon is called the spin Seebeck effect (SSE);\cite{Uchida2010,Uchida2010a}
for information on recent investigations,
see introduction of Ref.\cite{Adachi2018}
As the reciprocal phenomenon of the SSE,
the spin Peltier effect also has been investigated.\cite{Flipse2014,Daimon2016,Ohnuma2017}
On the other hand,
it is possible to use electric field as the driving force if a metal is attached to magnetic insulators,
due to the spin Hall effects\cite{Hirsch1999,Zhang2000,Adachi2013a}
or magnon--drag effects via electron--magnon interaction.\cite{Miura2012a,Ohnuma2017,Tang2018}
We now consider magnonic devices driven by a temperature gradient.
In 2013, Ren and Zhu proposed spin Seebeck diodes
by using two magnetic insulators separated by a nonmagnetic thin layer acting as a tunneling barrier;\cite{Ren2013}
hereinafter,
we call this multilayer structure a magnonic spin tunneling junction (MSTJ), following Ren and Zhu
(alternatively, this structure is called ferromagnetic insulating junction\cite{Nakata2018}).
It works as a diode for tunneling the spin current driven by the temperature difference between the two magnetic insulators,
and the spin Seebeck diode effect originates from
a magnon--magnon interaction that induces, in each magnet, different temperature dependences for the magnon density of states.
Here, it was assumed that the two magnetization directions are the same in both the magnetic insulators,
and the magnitude of the tunneling spin current was estimated as a function of temperature by using the typical parameters of YIG.
From this work, it can be seen that MSTJ has an additional degree of freedom, i.e.,
a relative angle between the left and right magnetization directions ($\bm n_\mathrm{L}$ and $\bm n_\mathrm{R}$, respectively) as shown in Fig. \ref{fig:fig1}.
The simplest way to control the magnetization directions is to apply an external magnetic field
to the MSTJs fabricated from magnetic materials with different anisotropy in the left and right magnets.

Here,
we investigate the thermal spin transport in MSTJs
and aim to reveal the relative angle dependence
between the left (L) and right (R) magnetization directions.
For this purpose, we theoretically describe the spin current in the presence of temperature difference, on the basis of a Heisenberg-type model
.
In addition to the diode property for the thermal tunneling spin current, which was already shown by Ren and Zhu,
we found that the spin current can be turned on/off by controlling the parallel/anti parallel condition.

Let us consider an MSTJ described by
\begin{align}
\HM&:=-\frac{1}{2}\sum_{i,j\in\gamma}J_{ij}
\hat{\bm S}_i
\cdot
\hat{\bm S}_j
-\sum_{i\in\gamma}\hat{\bm S}_i\cdot \bm H_\mathrm{eff}^\gamma,\\
\HI&:=-\sum_{i\in\mathrm{L},l\in\mathrm{R}}V_{il}\hat{\bm S}_i\cdot\hat{\bm S}_l,
\end{align}
where $\HM$ is the Hamiltonian of the spins within the $\gamma$--side magnetic insulator ($\gamma=\mathrm{L}$ or $\mathrm{R}$),
$\hat{\bm S}_i$ is the spin operator at a site $i$,
$J_{ij}=J_{ji}$ is an exchange coupling constant,
$\bm H_\mathrm{eff}^\gamma$ is a classical vector that denotes a homogeneous effective exchange field within the $\gamma$--side magnetic insulator
whose direction is parallel to $\bm n_\gamma$,
and $\HI$ is the tunnel Hamiltonian represented by a weak exchange coupling constant $V_{il}=V_{li}$
between site $i\in \mathrm{L}$ and site $l\in\mathrm{R}$ spins.
The equation of motion for the left--side spin operator is given by
$
\sum_{i\in\mathrm{L}}\dot{\hat{\bm S}}_i
=
\sum_{i\in\mathrm{L}}\hat{\bm S}_i\times\bm H_\mathrm{eff}^\mathrm{L}
-
\hat{\bm I}_\mathrm{s}
$,
where $\dot{\hat{O}}:=[\hat{O},\hat{\mathcal{H}}_\mathrm{L}+\hat{\mathcal{H}}_\mathrm{R}+\HI]/(\I\hbar)$ for an operator $\hat{O}$.
The first term represents the precession within the left--side magnetic insulator,
and the second term is the spin current operator passing from the left magnetic insulator to the right magnetic insulator, which defined by
\begin{align}
\hat{\bm I}_\mathrm{s}:=-\sum_{i\in\mathrm{L},l\in\mathrm{R}}V_{il}\hat{\bm S}_i\times\hat{\bm S}_l.
\end{align}
It should be noted that one can easily confirm the symmetry:
$
\sum_{i\in\mathrm{R}}\dot{\hat{\bm S}}_i
=
\sum_{i\in\mathrm{R}}\hat{\bm S}_i\times\bm H_\mathrm{eff}^\mathrm{R}
+
\hat{\bm I}_\mathrm{s}$.
Considering the inverse temperatures in the left and right magnetic insulators
as $\beta_\mathrm{L}$ and $\beta_\mathrm{R}$, respectively,
and assuming that the insulators form a contact via $\HI$,
the spin current $\bm I_\mathrm{s}$ can be calculated by using Zubarev's method.\cite{Zubarev1974}
In this method,
one can obtain up to the second--order calculation with respect to $V_{il}$ as
\begin{align}
\bm I_\mathrm{s}
=
-
\sum_{i\in\mathrm{L},l\in\mathrm{R}}V_{il}\braket{\hat{\bm S}_i}\times\braket{\hat{\bm S}_l}
+
\frac{1}{\I\hbar}
\lim_{\delta\to 0^+}
\int_0^\infty\D t
\E^{-\delta t}
\braket{
[
\hat{\bm I}_\mathrm{s}(t)
,\HI]
}
,
\label{eq:IS}
\end{align}
where
the statistical average is given by $\braket{\hat{O}}:=\Tr\E^{-\beta_\mathrm{L}\HL-\beta_\mathrm{R}\HR}\hat{O}/
\Tr\E^{-\beta_\mathrm{L}\HL-\beta_\mathrm{R}\HR}$,
and the time--dependent operator is denoted by the Heisenberg representation in $\HL+\HR$.
The first term is oriented along $\bm n_\mathrm{L}\times\bm n_\mathrm{R}$ because $\braket{\bm S_i}\propto\bm n_\gamma$ for $i\in\gamma$,
and
it does not vanish even in the equilibrium condition ($\beta_\mathrm{L}=\beta_\mathrm{R}$).
This term corresponds to a spin--system version of the conventional field--like spin torque
in a magnetic tunnel junction
consisting of ferromagnetic \textit{electrodes}.\cite{Tang2010,Miura2011a,Miura2012}
To evaluate the correlation function in the second term in Eq. (\ref{eq:IS}),
we represent the spin operator in terms of the local spin axis as
$
\hat{\bm S}_i=:\bm n_\gamma\hat S_i^z + \frac{1}{2}[(\bm n_\gamma^x+\I\bm n_\gamma^y)\hat S_i^- + (\bm n_\gamma^x-\I\bm n_\gamma^y)\hat S_i^+]
$
for $i\in\gamma$
where $\bm n_\gamma^x$ and $\bm n_\gamma^y$ are arbitrary unit vectors satisfying $\bm n_\gamma^x\times\bm n_\gamma^y=\bm n_\gamma$,
and the eigen values of $S_i^z$ are given by $-S_\gamma,-S_\gamma+\hbar,\ldots,S_\gamma$ in terms of a spin quantum number $S_\gamma>0$.

First, we discuss the collinear cases.
For $\bm n:=\bm n_\mathrm{L}=\bm n_\mathrm{R}$,
we have $\bm I_\mathrm{s}=I_\mathrm{s}^\para\bm n$ where
\begin{align}
I_\mathrm{s}^\para&:=
-
2\pi\hbar
\int_{-\infty}^\infty\D\omega
\left[
N_\mathrm{L}(\omega)
-
N_\mathrm{R}(\omega)
\right]
\Lambda^\para(\omega,\omega)
,
\\
\Lambda^\para(\omega,\omega')
&:=
S_\mathrm{L}S_\mathrm{R}
\sum_{i,j\in\mathrm{L}}
\sum_{l,m\in\mathrm{R}}
V_{il}
V_{jm}
\rho_{ij}^{+-}(\omega)
\rho_{ml}^{+-}(\omega')
,
\label{eq:Ren}
\end{align}
and $N_\gamma(\omega):=(\E^{\beta_\gamma\hbar\omega}-1)^{-1}$, and we introduce a spectral function on the junction interface defined by
\begin{align}
\rho_{ij}^{\alpha\beta}(\omega):=\frac{1}{2\pi}\int_{-\infty}^\infty\D t\E^{\I\omega t}
\frac{\braket{[\varDelta\hat{S}_i^\alpha(t),\varDelta\hat{S}_j^\beta]}}{2\hbar S_\gamma}
\quad\mathrm{for}\quad i,j\in\gamma
,
\end{align}
and $\varDelta\hat S_i^\alpha(t):=\hat S_i^\alpha(t)-\braket{\hat S_i^\alpha}$.
As seen from the Holstein--Primakoff transformation,\cite{Holstein1940}
$\rho_{ij}^{+-}(\omega)$ can be approximated by a magnon spectral function on the interface as $\rho_{ij}^{+-}(\omega)\simeq D_{ij}(\omega)$;
thus, we can confirm that
$I_\mathrm{s}^\para$ is equivalent to Ren and Zhu's result.\cite{Ren2013}
In another collinear case ($\bm n=\bm n_\mathrm{L}=-\bm n_\mathrm{R}$),
the spin current is given by
$\bm I_\mathrm{s}=I_\mathrm{s}^\apara\bm n$ where
\begin{align}
I_\mathrm{s}^\apara&:=
2\pi\hbar
\int_{-\infty}^\infty\D\omega
\left[
N_\mathrm{L}(\omega)
-
N_\mathrm{R}(\omega)
\right]
\Lambda^\apara(\omega,\omega)
,
\\
\Lambda^\apara(\omega,\omega')
&:=
S_\mathrm{L}S_\mathrm{R}
\sum_{i,j\in\mathrm{L}}
\sum_{l,m\in\mathrm{R}}
V_{il}
V_{jm}
\rho_{ij}^{+-}(\omega)
\rho_{lm}^{+-}(-\omega')
.
\end{align}
Here, assuming that the ferromagnetic magnon eigen frequency is positive in both the magnets,
the magnon spectral function has a finite value only for $\omega>0$.
Thus, we can conclude that the spin injection is almost perfectly blocked ($I_\mathrm{s}^\apara\simeq 0$) regardless of the sign of $V_{il}$,
because $\rho_{ij}^{+-}(\omega)\rho_{lm}^{+-}(-\omega)\simeq D_{ij}(\omega)D_{lm}(-\omega)\simeq 0$ for any frequency.
As a result,
it is possible to switch the tunnel spin current with high efficiency by controlling the magnetization direction of either the left or right magnetic insulator.

For an arbitrary relative angle,
we can represent the spin current
as
$
\bm I_\mathrm{s}
=
\bm I_\mathrm{s}^\mathrm{field}
+
\bm I_\mathrm{s}^\mathrm{neq}
$ where
\begin{align}
\bm I_\mathrm{s}^\mathrm{field}
&:=
-\sum_{i\in\mathrm{L}}
\sum_{l\in\mathrm{R}}
V_{il}
\bm S_i
\times
\bm S_l
+
\bm n_\mathrm{L}\times\bm n_\mathrm{R}
\left[
I_s'
+
(\bm n_\mathrm{L}\cdot\bm n_\mathrm{R})I_s''
\right]
,
\\
\bm I_\mathrm{s}^\mathrm{neq}
&:=
I_\mathrm{s}^1\bm n_\mathrm{L}\times\bm n_\mathrm{R}\times\bm n_\mathrm{L}
+I_\mathrm{s}^2\bm n_\mathrm{R}\times\bm n_\mathrm{L}\times\bm n_\mathrm{R}
\notag\\
&+I_\mathrm{s}^\para\frac{(1+\bm n_\mathrm{L}\cdot\bm n_\mathrm{R})(\bm n_\mathrm{L}+\bm n_\mathrm{R})}{4}
+I_\mathrm{s}^\apara\frac{(1-\bm n_\mathrm{L}\cdot\bm n_\mathrm{R})(\bm n_\mathrm{L}-\bm n_\mathrm{R})}{4}.
\end{align}
In addition, we introduce the statistically averaged $\hat{\bm S}_i$
on the basis of Zubarev's method up to the first--order $V_{il}$:
\begin{align}
\bm S_{i}
&:=
\braket{\hat{\bm S}_i}
+
S_\gamma
\dashint_{-\infty}^\infty\frac{\D\omega}{\omega}
\sum_{j\in\gamma}
\sum_{m\in\bar\gamma}
V_{jm}
\braket{\hat S_m^z}
\Re
\biggl\{
\bm n_{\bar\gamma}
\rho_{ij}^{+-}(\omega)
\notag\\
&+
\bm n_\gamma(\bm n_\mathrm{L}\cdot\bm n_\mathrm{R})
\left[
2\rho_{ij}^{zz}(\omega)
-
\rho_{ij}^{+-}(\omega)
\right]
\biggr\}
+\mathcal{O}(V_{il}{}^2)
\quad\mathrm{for}\quad i\in\gamma.
\label{eq:S}
\end{align}
Here, the symbol $\dashint$ indicates taking a principal value,
and $\bar{\mathrm{L}}:=\mathrm{R}$ ($\bar{\mathrm{R}}:=\mathrm{L}$).
The other parts are defined by
\begin{align}
I_\mathrm{s}'&:=
\hbar\dashint_{-\infty}^\infty\D\omega\D\omega'
\frac{N_\mathrm{L}(\omega)-N_\mathrm{R}(\omega')}{\omega-\omega'}
\frac{
\Lambda^\para(\omega,\omega')
+
\Lambda^\apara(\omega,\omega')
}{2}
,\\
I_\mathrm{s}''&:=
\hbar\dashint_{-\infty}^\infty\D\omega\D\omega'
\frac{N_\mathrm{L}(\omega)-N_\mathrm{R}(\omega')}{\omega-\omega'}
\Biggl(
\frac{
\Lambda^\para(\omega,\omega')
-
\Lambda^\apara(\omega,\omega')
}{2}
\notag\\
&-2\Lambda^1(\omega,\omega')
-2\Lambda^2(\omega,\omega')
+4\Lambda^3(\omega,\omega')
\Biggr)
,\\
I_\mathrm{s}^i&:=
-2\pi\hbar
\int_{-\infty}^\infty\D\omega
[N_\mathrm{L}(\omega)-N_\mathrm{R}(\omega)]
\Lambda^i(\omega,\omega),
\end{align}
and
\begin{align}
\Lambda^1(\omega,\omega')
&:=
S_\mathrm{L}S_\mathrm{R}
\sum_{i,j\in\mathrm{L}}
\sum_{l,m\in\mathrm{R}}
V_{il}
V_{jm}
\rho_{ij}^{zz}(\omega)
\rho_{ml}^{+-}(\omega')
,\\
\Lambda^2(\omega,\omega')
&:=
S_\mathrm{L}S_\mathrm{R}
\sum_{i,j\in\mathrm{L}}
\sum_{l,m\in\mathrm{R}}
V_{il}
V_{jm}
\rho_{ij}^{+-}(\omega)
\rho_{ml}^{zz}(\omega')
,\\
\Lambda^3(\omega,\omega')
&:=
S_\mathrm{L}S_\mathrm{R}
\sum_{i,j\in\mathrm{L}}
\sum_{l,m\in\mathrm{R}}
V_{il}
V_{jm}
\rho_{ij}^{zz}(\omega)
\rho_{ml}^{zz}(\omega')
.
\end{align}
The field--like term $\bm I_s^\mathrm{field}$ is oriented along $\bm n_\mathrm{L}\times\bm n_\mathrm{R}$
and does not vanish in the equilibrium condition.
On the other hand,
$\bm I_\mathrm{s}^\mathrm{neq}$
exists only in the non-equilibrium condition $\beta_\mathrm{L}\neq\beta_\mathrm{R}$,
and further, they are represented by a single frequency integration involving the product of the two spectral functions at the same frequency.
This suggests that they are the dissipative spin currents induced by the SSE.
In the magnon representation,
we can evaluate the spectral functions as
$\rho_{ij}^{+-}(\omega)\simeq D_{ij}(\omega)$
and
$\rho_{ij}^{zz}(\omega)\simeq\frac{\hbar}{2S_\gamma}\int_{-\infty}^\infty\D\omega'\left[
N_\gamma(\omega'-\omega)
-
N_\gamma(\omega')
\right]
D_{ij}(\omega')
D_{ji}(\omega'-\omega)
$;
thus, it can be seen that
the leading term in $\bm I_\mathrm{s}^\mathrm{neq}$ is given only by $I_\mathrm{s}^\para$ under the low temperature condition $\hbar N_\gamma(\omega)/S_\gamma\ll 1$.

In summary,
we have microscopically described the spin current induced by the SSE, which passes through the MSTJ
with arbitrary relative angles between the left and right magnetization directions.
Thus,
we have shown that it is possible to realize spin current switching for the spin Seebeck diode proposed by Ren and Xhu.\cite{Ren2013}
Furthermore, this switch potentially functions with high efficiency
because the blocking effect, $I_\mathrm{s}^\apara\simeq 0$,
is robust for junction interface conditions in the sense that it does not depend on the sign of the exchange coupling between
the magnetic insulators.

\begin{acknowledgment}
This work was supported by JSPS KAKENHI Grant No. 16K06702 and 17K14800 in Japan.
\end{acknowledgment}

\bibliographystyle{jpsj}
\bibliography{library}

\begin{thebibliography}{10}

\bibitem{Chumak2014}
A.~V. Chumak, A.~A. Serga, and B.~Hillebrands: Nat. Commun. {\bfseries 5}
  (2014) 4700.

\bibitem{Schneider2008}
T.~Schneider, A.~A. Serga, B.~Leven, B.~Hillebrands, R.~L. Stamps, and M.~P.
  Kostylev: Appl. Phys. Lett. {\bfseries 92} (2008) 022505.

\bibitem{Kajiwara2010}
Y.~Kajiwara, K.~Harii, S.~Takahashi, J.~Ohe, K.~Uchida, M.~Mizuguchi,
  H.~Umezawa, H.~Kawai, K.~Ando, K.~Takanashi, S.~Maekawa, and E.~Saitoh:
  Nature {\bfseries 464} (2010) 262.

\bibitem{Zutic2004}
I.~{\v{Z}}uti{\'{c}}, J.~Fabian, and S.~D. Sarma: Rev. Mod. Phys. {\bfseries
  76} (2004) 323.

\bibitem{Lenk2011}
B.~Lenk, H.~Ulrichs, F.~Garbs, and M.~M{\"{u}}nzenberg: Phys. Rep. {\bfseries
  507} (2011) 107.

\bibitem{Bauer2012}
G.~E. Bauer, E.~Saitoh, and B.~J. {Van Wees}: Nat. Mater. {\bfseries 11} (2012)
  391.

\bibitem{Uchida2010}
K.~Uchida, J.~Xiao, H.~Adachi, J.~Ohe, S.~Takahashi, J.~Ieda, T.~Ota,
  Y.~Kajiwara, H.~Umezawa, H.~Kawai, G.~E. Bauer, S.~Maekawa, and E.~Saitoh:
  Nat. Mater. {\bfseries 9} (2010) 894.

\bibitem{Uchida2010a}
K.~Uchida, T.~Nonaka, T.~Ota, and E.~Saitoh: Appl. Phys. Lett. {\bfseries 97}
  (2010) 262504.

\bibitem{Adachi2018}
H.~Adachi, Y.~Yamamoto, and M.~Ichioka: J. Phys. D. Appl. Phys. {\bfseries 51}
  (2018) 144001.

\bibitem{Flipse2014}
J.~Flipse, F.~K. Dejene, D.~Wagenaar, G.~E. Bauer, J.~B. Youssef, and B.~J.
  {Van Wees}: Phys. Rev. Lett. {\bfseries 113} (2014) 027601.

\bibitem{Daimon2016}
S.~Daimon, R.~Iguchi, T.~Hioki, E.~Saitoh, and {Ken-ichi Uchida}: Nat. Commun.
  {\bfseries 7} (2016) 13754.

\bibitem{Ohnuma2017}
Y.~Ohnuma, M.~Matsuo, and S.~Maekawa: Phys. Rev. B {\bfseries 96} (2017)
  134412.

\bibitem{Hirsch1999}
J.~E. Hirsch: Phys. Rev. Lett. {\bfseries 83} (1999) 1834.

\bibitem{Zhang2000}
S.~Zhang: Phys. Rev. Lett. {\bfseries 85} (2000) 393.

\bibitem{Adachi2013a}
H.~Adachi, K.~I. Uchida, E.~Saitoh, and S.~Maekawa: Reports Prog. Phys.
  {\bfseries 76} (2013) 036501.

\bibitem{Miura2012a}
D.~Miura and A.~Sakuma: J. Phys. Soc. Japan {\bfseries 81} (2012) 113602.

\bibitem{Tang2018}
G.~Tang, X.~Chen, J.~Ren, and J.~Wang: Phys. Rev. B {\bfseries 97} (2018)
  081407(R).

\bibitem{Ren2013}
J.~Ren and J.~X. Zhu: Phys. Rev. B {\bfseries 88} (2013) 094427.

\bibitem{Nakata2018}
K.~Nakata, Y.~Ohnuma, and M.~Matsuo: Phys. Rev. B {\bfseries 98} (2018) 094430.

\bibitem{Zubarev1974}
D.~N. Zubarev: {\em {Nonequilibrium Statistical Thermodynamics}} (Consultants
  Bureau, New York, 1974).

\bibitem{Tang2010}
Y.-H. Tang, N.~Kioussis, A.~Kalitsov, W.~H. Butler, and R.~Car: J. Phys. Conf.
  Ser. {\bfseries 200} (2010) 062033.

\bibitem{Miura2011a}
D.~Miura and A.~Sakuma: J. Appl. Phys. {\bfseries 109} (2011) 07C909.

\bibitem{Miura2012}
D.~Miura and A.~Sakuma: J. Phys. Soc. Japan {\bfseries 81} (2012) 054709.

\bibitem{Holstein1940}
T.~Holstein and H.~Primakoff: Phys. Rev. {\bfseries 58} (1940) 1098.

\end{thebibliography}

\end{document}